\documentclass{ringb99}
\usepackage{graphics}

\begin{document}
\title{A Comparison of the Extra Nuclear X-ray and Radio Features in M87}
\author{D. E. Harris\inst{1}, F. Owen\inst{2}, J. A. Biretta\inst{3}, \and W. Junor\inst{4}}  
\institute{Smithsonian Astrophysical Observatory,
 60 Garden St., Cambridge, MA 02138 USA; harris@cfa.harvard.edu
\and National Radio Astronomy Obs.
 P.O. Box 0, Socorro, NM 87801, USA; fowen@nrao.edu 
\and  Space Telescope Science Institute, 
 3700 San Martin Drive, Baltimore, MD 21218, USA; biretta@stsci.edu
\and Dept. of Physics and Astronomy,
  University of New Mexico,
  800 Yale Blvd., NE, Albuquerque, NM 87131, USA; bjunor@astro.phys.unm.edu}

\authorrunning{Harris, Owen, Biretta, \& Junor}
\titlerunning{X-ray and Radio Features in M87}
\maketitle

\begin{abstract}

ROSAT High Resolution Imager (HRI) data from eight observations have
been co-added to obtain an effective exposure of 230 ksec.  We have
identified a number of features and regions with excess X-ray
brightness over that from a circularly symmetric model of the well
known hot gas component.  A prominent `spur' extends 4$^{\prime}$ from
the vicinity of knot A towards the south-west.  The brightness to the
south and east of this spur is significantly higher than that to the
north and west.  Excess brightness is also found to the East of the
nucleus, with a local maximum centered on the eastern radio lobe
3$^{\prime}$ from the core.

There are two well known relationships between radio and x-ray
emission for radio galaxies in clusters: coincidence of emissions
because the X-rays come from inverse Compton or synchrotron processes;
and anti-coincidence caused by exclusion of hot gas from radio
entities.  We present a radio/X-ray comparison to determine if either
of these relationships can be isolated in M87.  The greatest
obstacle we face is the unknown projection which affects both bands.

\end{abstract}

\section{Introduction}

Eight ROSAT/HRI observations of M87 were made between 1992Jun and
1998Jan to study the X-ray structure and variability of the core and
jet (Harris, Biretta, and Junor, 1997 and 1998a).  To study the large
scale X-ray features of M87, we have used these data to make an image
with effective exposure of 230 ksec.  A preliminary analysis based
on the data then available was presented at the Ringberg Workshop on
M87 held in 1997Sep (Harris, Biretta, and Junor 1998b).

The radio map of M87 (see Owen, this volume) appears to indicate an
exceedingly complex structure, for which it is difficult to make
meaningful measurements of isolated features.  Generally, we need to
define volumes and measure their emission properties.  Once you leave
the inner (brightest) lobes and jet, this is not easily done.  Even
for definable filaments and other features, the surface brightness is
often a sum of emissivities from various entities along the line of
sight.  The X-ray map suffers from the same problem although generally
there is less fine structure than in the radio (at similar
resolutions).  An additional complexity for the X-ray analysis is that
we cannot be certain that the very large scale X-ray distribution from
the hot gas of the Virgo cluster is circularly symmetric.

\section{M87 as a Wide Angle Tailed Radio Galaxy}

One possible interpretation of the radio map is that we are viewing a
Wide Angle Tailed (WAT) radio galaxy from an angle such that a large
fraction of the radio source is well beyond (or in front of) the
central part of the galaxy.  Thus the observed radio brightness at any
given location will usually be the sum of contributions from several
different emitting volumes.  There are at least three supporting
arguments for this hypothesis, although none is conclusive.

Without invoking substantial projection effects, it is difficult to
maintain continuity between various radio features.  In the bright inner
region, we see the primary jet bending around to the south, yet the larger
scale emission which one might expect to connect to the bright features,
is found due West.  The same situation occurs on the East side of the
source.

Without projection, many features display rather strong curvature.
Besides the inner lobes, projection effects may be the cause for the
apparent curvature in the eastern double ring and in the `L' filament
at the south-east edge of the source.  The double ring itself is
difficult to understand if it were completely in the plane of the sky.
The radio galaxy $0053-016$ in Abell 119 (Feretti et al. 1999; see
also Govoni et al. in thses proceedings) shows a helical structure
which, if viewed along the arm of the tail, could project to something
like the apparent structure of the double ring.

Pressure balance between the ambient gas and the non-thermal pressure
within radio features is expected for most of the radio structures
except those within the very bright central region.  We have compared
non-thermal pressures for two low brightness regions close to the
center with the expected thermal pressure derived from the work of
Nulsen and B\"{o}hringer (1995).  As shown in Table~\ref{tab:nt_pres},
the thermal pressure exceeds the minimum non-thermal pressures for
both regions.  We integrated the radio spectrum between 10$^7$ and
10$^{10}$ Hz and used spectral index values of 0.8 and 0.9 (from a
spectral index map between 74 and 327 MHz provided by N. Kassim).  The
entries for k=0 represent the minimum pressure for the case of the
filling factor, $\phi$=1 and no contribution to the particle energy
density from relativistic protons (k=0).  The entries with k=100
correspond to either protons having 100 times the energy density of
the relativistic electrons or $\phi$=0.01.  The last column indicates
the radial distance required to reach pressure balance between the
thermal gas and the non-thermal radio feature.  In all cases, these
distances are greater than the projected distance from the center.

\begin{table}
      \caption{Non-thermal pressure estimates.}
         \label{tab:nt_pres}
      \[
         \begin{array}{lclcc}
            \hline
            \noalign{\smallskip}
            Feature^{\rm a}    &  B^{\rm b}  &  P(nt)^{\rm c} &
            n_e(bal)^{\rm d}  & R^{\rm e}\\
	    & (\mu G) & (dyne~cm^{-2}) & (10^{-3}cm^{-3})  & (arcmin)\\
            \noalign{\smallskip}
            \hline
            \noalign{\smallskip}
            cube (k=0)   &  6  &  3\times10^{-12}  &  0.7  & 39   \\
            cube (k=100) & 22  &  4\times10^{-11}  &  9.7  &  4   \\
	    cyl  (k=0)   &  7  &  5\times10^{-12}  &  1.1  & 26   \\
	    cyl  (k=100) & 28  &  7\times10^{-11}  & 15    &  2.5 \\ 
            \noalign{\smallskip}
             \hline
         \end{array}
      \]
\begin{list}{}{}
\item[$^{\rm a}$] The cube is located 90$^{\prime\prime}$ SW of the
nucleus (the head of the `cobra', RA$\approx$12h30m43s and
DEC$\approx$12$^{\circ}$23$^{\prime}$).  The cylinder is
72$^{\prime\prime}$ due east of the nucleus, a segment of the `eastern
arm' at RA$\approx$12h30m54s,
DEC$\approx$12$^{\circ}$23$^{\prime}$45$^{\prime\prime}$. $k$ is the
factor for energy density from relativistic protons.  k=100 is
equivalent to a filling factor of $\phi$=0.01.  The thermal pressure
expected at the radial distances of the cube and cylinder are 1.05 and
1.2 $\times$~10$^{-10}$ dyne cm$^{-2}$.

\item[$^{\rm b}$] B is the magnetic field strength for equipartition.

\item[$^{\rm c}$] P(nt) is the non-thermal pressure.

\item[$^{\rm d}$] n$_e$(bal) is the thermal electron density required to
balance the non-thermal pressure for a temperature of 2 keV.

\item[$^{\rm e}$] R is the radial distance at which the model density
equals n$_e$(bal).

\end{list}
   \end{table}

\begin{figure}
\label{fig:m87all}
 \resizebox{\hsize}{!}{\includegraphics{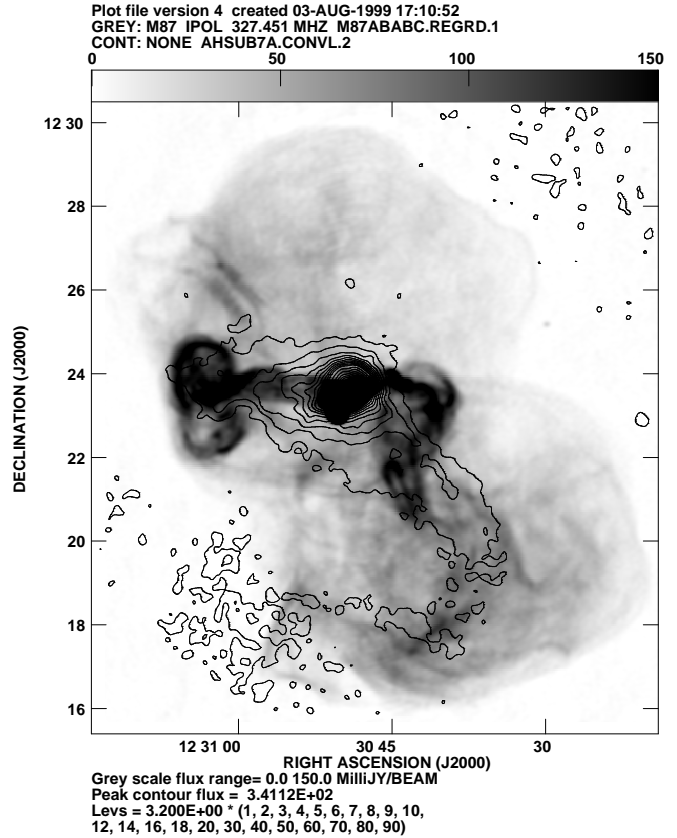}}
\caption[M87 all]{An overlay of the X-ray map (contours) on the VLA
327 MHz map (grey scale).  The radio map has a
beamwidth=7.8$^{\prime\prime}\times$6.5$^{\prime\prime}$.  The X-ray
map has been shifted to align the radio and X-ray cores.  A Gaussian
of FWHM=10$^{\prime\prime}$ was used to smooth the X-ray map.  The
southwest spur is delineated by the second (narrow) contour.  More
detail can be seen in the color version of this figure which is with
other color figures in this book.}
\end{figure}

There are various possible reasons for a relative motion between the
ICM and M87 which could produce a WAT structure.  Harris et
al. (1998b) suggested that the SW X-ray spur (see fig. 1) might be
caused by a shock between the ISM of M87 and the ICM disturbed by a
merger of the M86 subgroup with the M87 subgroup.  We note the
relative velocity of M87 and M86, $\sim 1500$ km s$^{-1}$, is
sufficient to shock $10^{7}$ K gas.  Other possibilities are that M87
might not be at the center of the potential well of the Virgo cluster
or that the cooling flow might be asymmetric if the kinetic energy
supplied by the movement of radio structures supplied heat to the
local gas in a non symmetric fashion.

\section{Data}

The data reduction for the present comparison of radio and X-ray maps
involved the following steps.

\begin{itemize}

      \item Five of the eight X-ray observations were corrected for
the bug in the standard processing which used incorrect aspect times
(see the documents available at the anonymous ftp server:
sao-ftp.harvard.edu; cd to pub/rosat/aspfix).  The other three
observations were made after 1997Jan when this bug had essentially no
effect on image quality.

      \item The centroid of the core emission was measured for each
observation and each map was then shifted to a common position before
stacking.

      \item The radial profile of the total emission was measured in a
90$^{\circ}$ quadrant towards the north west, centered on the peak of the
core emission.

      \item On the basis of this profile, a circularly symmetric King
model was constructed to represent the bulk of the cluster emission.
The model was then subtracted from the data and the residual was
smoothed with Gaussians of various widths.  Note that this was more of
a procedural process than a true modelling of the emission since the
profile does not match a simple King model.  It does, however,
emphasize the contrast between residual features and the remaining
background.

      \item Finally, the radio map was precessed to J2000 and the
X-ray map was shifted so as to align the X-ray and radio core
emissions.

\end{itemize}

\section{Radio/X-ray Coincidence}

Coincidence between radio and X-ray emissions can be expected when
both arise from non-thermal processes: i.e. the X-ray emission is
either synchrotron or inverse Compton emission.  Although successful
models for synchrotron X-ray emission from knot A in the M87 jet have
been published (Biretta, Stern, and Harris 1991), the (presumably)
older and larger radio structures (fig. 1) under discussion here are
not expected to provide the environment to produce the very high
energy electrons (Lorentz energy factors $\gamma\approx$10$^7$)
required for X-ray synchrotron emission.  Inverse Compton emission
however, will be present, both from the 3K background photons
(`IC/3K') and from other photons such as star light.

There are only two regions where a general coincidence of radio and
X-ray emissions are found.  These are (a) in the southern part of the
radio source where a curved feature, the bottom of the (so-called)
`cobra' shows low brightness enhancements in both bands
(fig. 2); and (b) the `eastern radio arm' extending
due East from the core and ending in the region at the center of the
double radio rings (fig. 3).  Note however that even
here, the spatial agreement is not precise.

\begin{figure}
\label{fig:m87south}
 \resizebox{\hsize}{!}{\includegraphics{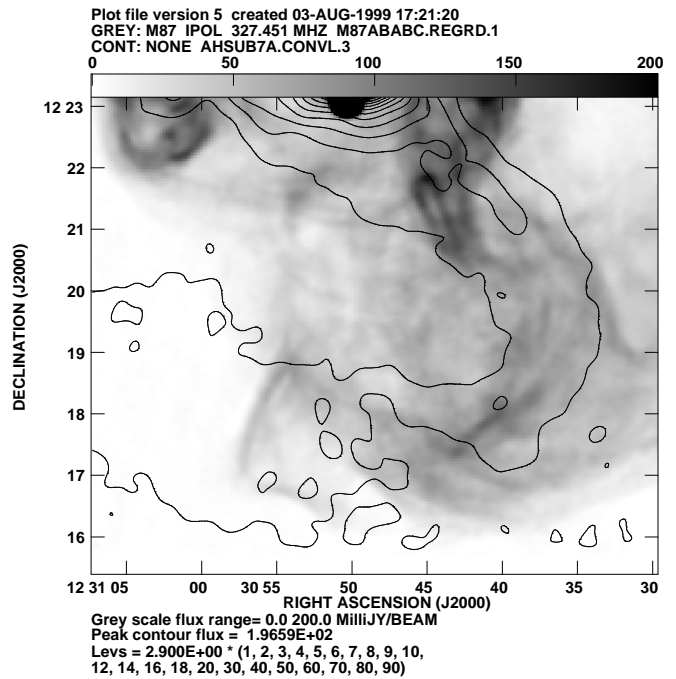}}
\caption[M87 south]{The southern portion of M87.  The grey scale is
the VLA map at 327 MHz and the X-ray map is shown by the contours.  It
has been smoothed by a Gaussian with FWHM=20$^{\prime\prime}$.  The
prominent feature curving to the south from the southwest spur is
termed ``the lower part of the cobra''.}
\end{figure}
  
\begin{figure}
 \label{fig:m87east}
 \resizebox{\hsize}{!}{\includegraphics{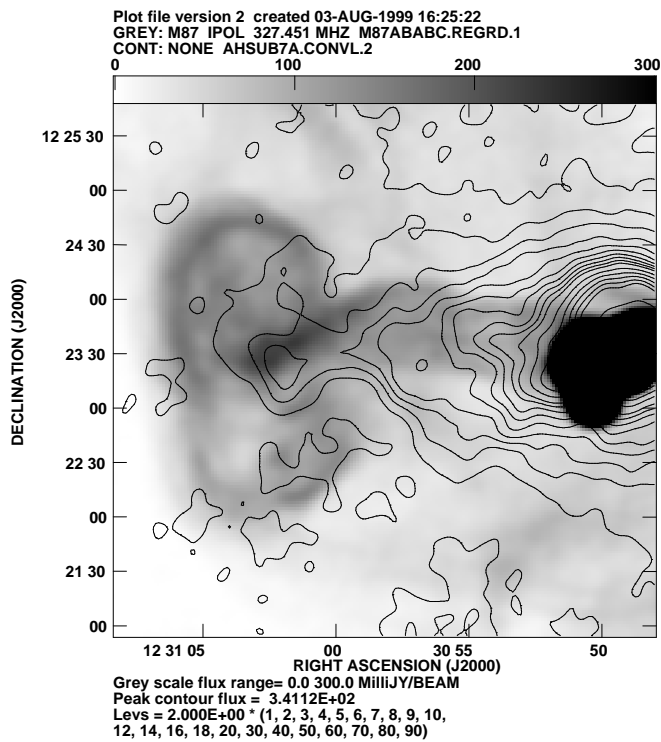}}
\caption[M87 east]{The Eastern part of M87.  The grey scale is from
VLA data at 327 MHz with a
7.8$^{\prime\prime}\times$6.5$^{\prime\prime}$ beamwidth.  The X-ray
contours are from a map smoothed with a 10$^{\prime\prime}$ Gaussian.
Note that whilst there is emission in both bands from the `eastern
arm' leading towards the double radio rings, there is no detailed
coincidence.  The brightest X-ray extension towards the east
corresponds to an area of low radio brightness and the X-ray peak at
the end of the arm does not align well with the peak radio
brightness.}
\end{figure}

Because most of the radio and residual X-ray emissions are not
co-spatial in a detailed sense, we conclude that the bulk of the
residual X-ray emission is likely to be thermal.  For example, the
general association of the radio and X-ray emission might arise from
heating of the ISM by the passage of the radio jets.  It has become
fashionable (at least at this meeting) to suggest that relativistic
particles do not necessarily occupy the same volumes defined by radio
emitting regions containing the strongest average magnetic field
strength.  Thus one could have IC/3K emission from regions essentially
devoid of radio emission.  While this is a valid physical scenario,
with our presently available technology, we have no methods to
demonstrate that any given X-ray brightness comes from IC processes
other than coincidence with known non-thermal emission at other bands
and the requisite calculations to demonstrate that the observed
intensity and implied field strengths are reasonable.  We also note
that many of the brighter radio features are devoid of spatially
coincident X-ray emission, which tends to argue against the IC
process.

\section{Radio/X-ray Anti-coincidence}

Anti-coincidence is expected for cases in which radio features exclude
the hot gas and thus diminish the X-ray surface brightness for
particular lines of sight.  Convincing examples of this behavior have
been found for Cygnus A (Carilli, Perley, and Harris, 1994) and for
NGC 1275 (B\"{o}hringer et al. 1993).  For simple models of lobe
expansion, the change in X-ray surface brightness between lines of
sight which intersect the lobe and those that traverse only
undisturbed ICM can be either positive or negative.  This is because
we expect to find a sheath of enhanced density around the inflating
lobe.  The precise behavior for any given situation depends on the
energy range covered by the telescope system and on the density,
temperature, and thickness of the sheath (Clarke, Harris, and Carilli,
1997).

For M87, most of the source outside the high brightness (inner) radio
lobes may not fulfill the simple conditions of an expanding lobe with
a well defined wall separating it from the ICM and the unknown
projection effects may mask the signatures of cavities in the gas.
However, there are two possibilities for this sort of effect.  The
residual map shows a slight discontinuity in brightness gradient along
the northern boundary of the radio source which may indicate a smaller
integrated emissivity when looking through the lobe (see fig. 4).  The
other effect is evidenced by two regions of slightly negative
brightness in the residual map.  These areas are identified in the
color figure by the single contour level with center
$\approx$~4$^{\prime}$ to the northeast of the core and a similar
feature located $\approx$~3$^{\prime}$ west of the core.  It is
difficult to evaluate the reality of these features because the
subtraction of the model is somewhat arbitrary since there is no
suitable pie segment in which to measure the true radial profile of
`undisturbed' gas.

\begin{figure}
 \label{fig:m87north}
\rotatebox{-90}{
 \resizebox{!}{\hsize}{\includegraphics{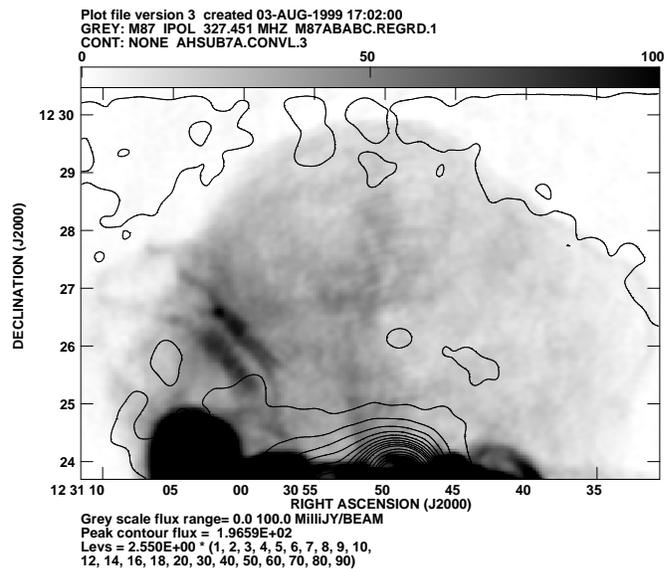}}}
\caption[M87north]{The northern part of M87.  The grey scale shows a
VLA map at 327 MHz with a
7.8$^{\prime\prime}\times$6.5$^{\prime\prime}$ beam.  The residual
X-ray map has been smoothed with a 20$^{\prime\prime}$ FWHM Gaussian.
The central part of this display is at a lower level than the upper
corners.  Note how the single contour roughly delineates the edge of
the radio source.}
\end{figure}

\section{Conclusions}

\begin{itemize}

      \item We suspect that projection effects confuse the
interpretation of both radio and X-ray features.  Harris et
al. (1998b) suggested that the SW spur might be caused by a bow shock
between the ICM and the ISM, and the resulting change in temperature
and density might explain the presence of knot A in the jet.  Since
the current evidence supports the notion that the spur is a thermal
feature, a convincing explanation for the spur will probably await the
spectral/spatial capabilities of Chandra and XMM.

      \item Many X-ray and radio features are located in the same
general region, but it appears that they do not actually occupy the
same volumes.

      \item There is some evidence for edge effects and cavities.

\end{itemize}

\begin{acknowledgements}
N. Kassim kindly provided us with an unpublished spectral index map.
WJ thanks the National Science Foundation for support under Grant
AST-980307.  The work at SAO was partially supported by NASA contract
5-99002.

\end{acknowledgements}

\end{document}